# Effects of Poly(vinylpyrrolidone) on the Dynamic Viscosity of Methane Hydrate Systems at High-Pressure Driving Forces: Investigation of Concentration, Molecular Weight, and Shear Rate

*Chong Yang Du, André Guerra, Adam McElligott, Milan Marić, Alejandro Rey, Phillip Servio\**

Department of Chemical Engineering, McGill University, Montreal, Quebec H3A 0C5, Canada

\*phillip.servio@mcgill.ca

## Abstract

The viscosity of methane hydrate slurries with poly(vinylpyrrolidone) (PVP) at 700 and 7000 ppm by weight, molecular weights of 40,000 (PVP40) and 360,000 (PVP360) Da, and shear rates of 400 and 80 s$^{-1}$, were measured in a high-pressure rheometer with pressures up to 30 MPag and compared to pure water systems. The additives successfully reduced the formation of high-viscosity slurries, but at low concentrations were incapable of delaying hydrate agglomeration at the late growth stage. The average relative time required for PVP40 solutions at 700 ppm to grow to 50 mPa·s was 1.9 times the water reference value, but only 1.2 times to reach 200 mPa·s. Improved inhibition was observed for the higher concentration and higher molecular weight sets, where the relative time to reach 50 mPa·s were 8.2 and 2.6 times the water reference value, respectively. While the additives demonstrated anti-nucleation properties and suppressed crystal growth initially, they accelerated the hydrate clusters agglomeration rate, and potentially weakened the hydrate mechanical properties.

# 1. Introduction

Under high pressure, water molecules can form crystals at temperatures much higher than 0 °C when stabilized by van der Waals interactions with dissolved gas molecules.[1, 2] Such compounds have a snow-like appearance and are commonly named clathrate hydrates or gas hydrates. Due to hydrogen bonding, water molecules can align into various cage structures which are physically supported by fitting guest species such as methane, ethane, and carbon dioxide.[1]

Therefore, when water accumulates in offshore oil and gas pipelines, the operating conditions under the seabed are often favourable for hydrate formation. Methane hydrates present a prominent concern for flow assurance in the energy sector, as their growth can block the pipes and result in major safety hazards, equipment damage, financial loss, and catastrophes for the environment.[3, 4]

The most frequently employed hydrate inhibitors are antifreezes such as methanol and ethylene glycol.[5, 6] Alcohols and glycols are categorized as thermodynamic inhibitors, which operate by shifting the hydrate forming conditions to a lower temperature and higher pressure.[1] Although effective, these antifreezes are typically required in significant concentrations in the system, ranging from 20 to 50 wt.%.[4, 7] To reduce operating and capital costs, much research has focussed on low dosage hydrate inhibitors (LDHIs).[8, 9]

Amongst LDHIs, the commercially available water-soluble polymer poly(vinylpyrrolidone) (PVP) is the most common. It is considered to be in the sub-category of kinetic hydrate inhibitors (KHIs), and is added at concentrations of 0.01 to 5 wt.% in the system.[4, 8, 10] It delays hydrate agglomeration in pipelines by disrupting both crystal nucleation and growth, preferably to times longer than the free water residence time. Its inhibitory properties rely on its cyclic amide group, which can engage in hydrogen bonding. By adsorbing onto the hydrate surface, the polymer reduces the diffusion of water and gas molecules into the hydrate phase. Although necessary, the ability to form hydrogen bonds is not the only requirement for KHIs. The increasingly hydrophilic poly(vinyl alcohol), for example, acts as a poor hydrate inhibitor compared to PVP.[10, 11]

KHIs' inhibitory performance is often measured in terms of the rate of moles of gas consumed by the hydrate-forming system. After the saturation and stochastic induction phases, hydrate former molecules such as methane or ethane are depleted linearly with time in the initial growth stage.[12] The addition of PVP in the aqueous phase, even at a small amount such as 700 ppmw, has proven to decrease methane consumption drastically at high-pressure driving forces, from 50% reduced gas intake to nearly zero growth.[11, 13]

Although the ability of PVP to reduce hydrate nucleation and growth rates is generally acknowledged in the field, contradicting claims on its performance have been reported. PVP has acted as both an inhibitor and promoter of hydrate formation.[10, 14, 15] Some studies even reported a greater amount of gas consumed in the presence of PVP compared to a water control.[9, 16] The results depended on a large number of variables such as subcooling, pressure, stir rate, gas and liquid phase composition, concentration, etc. Further investigation on PVP's hydrate inhibitory properties is required to gain deeper understanding of its working principle. This work sets out to investigate the effects of PVP on the dynamic viscosity of methane hydrate systems as measured by a high-pressure rheometer. This work considers the addition of PVP in different concentrations and molecular weights. Additionally, various shear rates are studied to establish the impact of mixing on hydrate growth and effectiveness of PVP.

## 2. Materials and Methods

2.1 Materials

The poly(vinylpyrrolidone) polymers PVP40 and PVP360 were purchased from Sigma-Aldrich, with number average molecular weights ($M_n$) of 40,000 and 360,000 Da, respectively. Several stock solutions, each weighing 100 grams, were prepared at concentrations of 700 and 7000 ppm by dissolving 0.07 and 0.7 g of PVP in reverse osmosis (RO) treated water. The RO water has a final conductivity of 10 µS and a maximum organic content of 10 ppb. The solutions were stirred by a magnetic stirrer for 15 minutes at 360 rpm at room temperature, then filtered using 0.22 µm PTFE filters (Fisher) before being loaded in the apparatus. The 100 g stock solutions were depleted or exchanged within 2 days to reduce the possibility of aggregates. Methane was supplied by MEGS Specialty Gases and Equipment Inc. at ultra high purity (99.99%).

2.2 Apparatus

The experimental setup employed in this report was slightly modified from that illustrated in the previous work of Guerra et al. (2022).[17] The rheological experiments were performed in an Anton Paar Modular Compact Rheometer 302 (MCR302), equipped with a pressure cell (PR400/XL) that had a maximum pressure rating of 40 MPag. The selected measuring system is the double-gap (DG) geometry DG35.12/PR/TI-A1, as it is considered to have the highest accuracy for measuring aqueous, low-viscosity solutions due to its high contact area with the sample. This geometry is composed of two concentric cylinders acting as the base cup, creating a gap where the test sample was loaded and where a hollow cylinder bob was inserted to induce shear (Fig. 1). The measurement device outputs data to the Anton Paar RheoCompass software, which monitors and controls the rheometer.

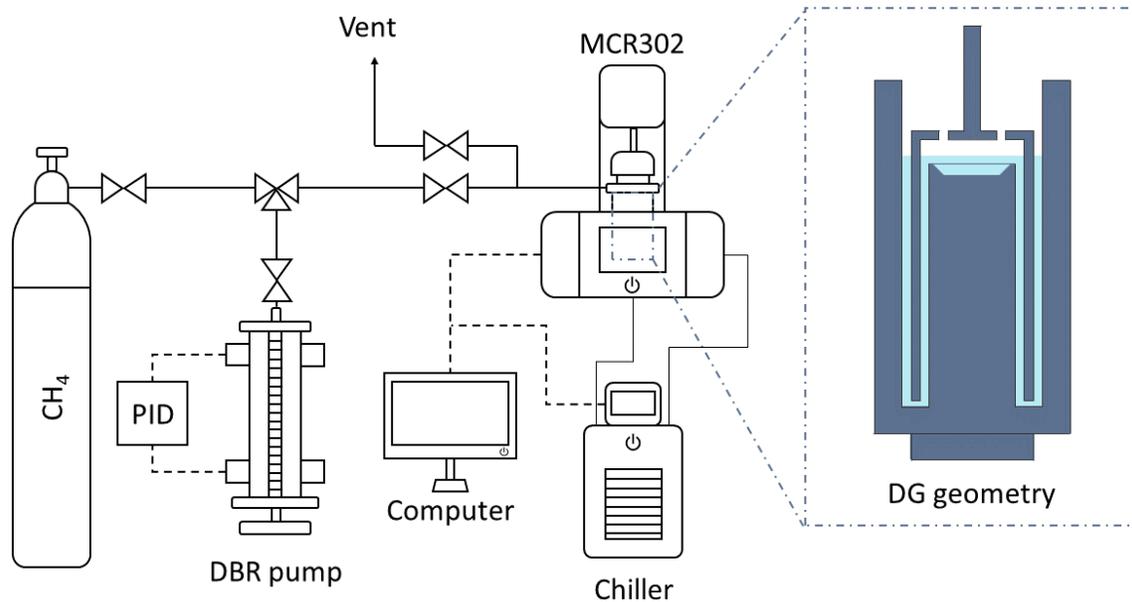

Fig. 1 Experimental setup for high-pressure rheological experiments.

A tee valve connects the pressure cell's gas inlet line to a methane gas cylinder, and to a Schlumberger DBR positive displacement pump. The system's pressure is controlled by the Schlumberger DBR Oil-Phase control software, and its temperature is controlled by a Julabo F32 chiller, whose refrigerant is a 50/50 mixture by volume of ethylene glycol and water. The pressure and temperature sensors of the pressure cell are a PA-23S/400bar transmitter and a PT100 thermometer, respectively.

2.3 Experimental procedure

With the pressure cell inlet valve closed, methane from the gas cylinder filled up the 500-cc piston of the DBR pump. Then, the tee valve, the gas regulator valves, and gas cylinder were closed to cut the gas flow between the cylinder and the pump for safety considerations. As the pressure and temperature controllers began operating, a 7.5 mL solution of PVP was filtered and loaded into the pressure cell. After being sealed, the pressure cell was purged at a pressure of 1 MPag five times for a period of 30 seconds each, to displace air in the system with methane.

As the temperature setpoint was reached (±0.10 °C), the inlet valve was opened to let methane flow from the pump to the pressure cell. The pressure controller would adjust and maintain the system pressure to ±0.30 MPag of the setpoint. Upon pressurization, the system temperature would first decrease due to gas expansion, then warm up because methane was stored at room temperature. The system might not have time to re-establish its temperature setpoint after pressurization.

As shown in the example in Fig. 2 (left), the temperature increased a second time at around 43 minutes, indicating rapid hydrate growth, which released energy and heated up the system. In the cases where this rapid hydrate growth occurred within a few minutes after pressurization (i.e., conditions with high-pressure driving forces such as 30 MPag or low temperatures such as 0 °C), the system temperature can remain slightly above the setpoint throughout the run. However, note that this second temperature spike does not indicate the end of the induction phase. As shown in Fig. 2 (right), the presence of PVP effectively slows down hydrate growth initially. Therefore, the nucleation and formation of low-viscosity hydrate slurries was not detected by the PT100. Rather, a spike in the dynamic viscosity graph (at around 13 minutes) marks the beginning of hydrate growth. A MATLAB code was used to process the CSV files recorded by RheoCompass and to output the time required for hydrates to grow to 3 (nucleation), 50, 100, 200, 500, 800, 1000, 1500, and 2000 mPa·s.

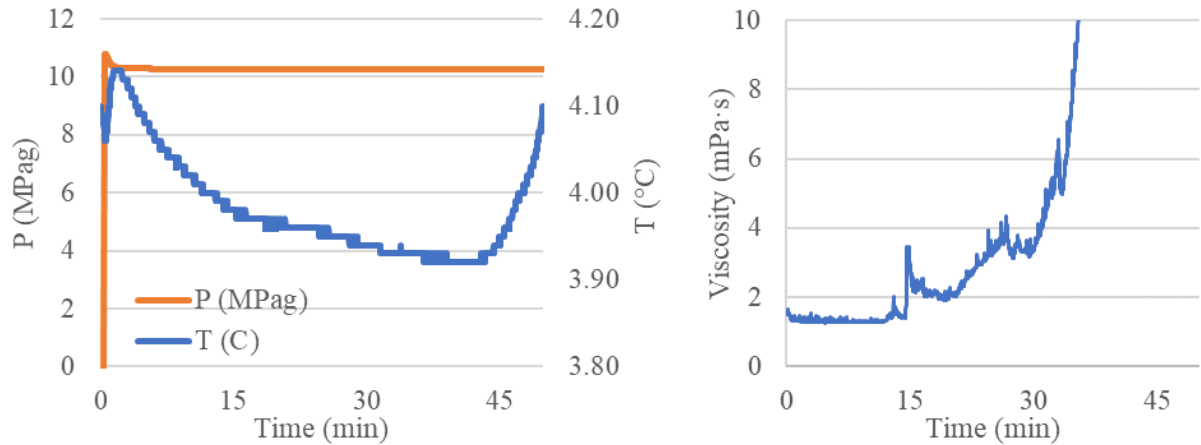

Fig. 2 Variables monitored during hydrate-forming runs. Left: temperature and pressure profiles and Right: dynamic viscosity profile (zoomed in) with respect to time

The pressure and temperature conditions of this experiment are the same as in previous works. For the purpose of comparing with a pure water-methane system at the same driving forces, results by Guerra et al. (2022)[17] were used as the reference system. For non-hydrate runs, the pressures were 0, 1, 2, 3, 4, and 5 MPag and for hydrate-forming runs, 10, 15, 20, 25, and 30 MPag. The experimental temperatures were 0, 2, 4, 6, 8, and 10 °C. The thermodynamic conditions at which hydrate formation was recorded in the MCR302 do not exactly correspond to the calculations by Carroll (2014).[1] Rather, they are limited by heat- and mass-transfer of the equipment, as well as the experimental timespan allowed for each run, which is given as 90 minutes to show any sign of growth.

In this report, the behavior of methane hydrates at 400 and 80 $s^{-1}$ was investigated. The reference water system has been tested under a 400 $s^{-1}$ shear rate, typical value for the measuring system to register accurate readings with low-viscosity, aqueous samples. A much lower shear rate of 80 $s^{-1}$ was also tested in the current study to verify the hypothesis of whether 400 $s^{-1}$ would inhibit the formation of critical hydrate clusters, as suggested in the previous reports.[18] Measurements were averaged over a full revolution (AOFR) with an adjustment time of 5 seconds for non-hydrate runs and 1 second for hydrate-forming runs. A torque limit of 115 mN·m was imposed to stop the measurement before high-viscosity hydrate slurries could solidify and damage the equipment. Three runs per condition was performed to verify the reproducibility of hydrate growth.

Table 1 Test conditions including shear rates, molecular weights, concentrations, temperatures, and pressures

| Shear rates ($s^{-1}$) | 400 | | | 80 | |
|---|---|---|---|---|---|
| $\bar{M}_{n,PVP}$ (kDa) | 40 | 40 | 360 | 0 | 40 |
| [PVP] (ppmw) | 700 | 7000 | 700 | 0 | 700 |
| $T_{exp}$ (°C) | 0, 2, 4, 6, 8, 10 | | | | |
| $P_{exp}$ (MPag) | 10, 15, 20, 25, 30 | | | 10, 15, 20 | |
| # of hydrate-forming runs | 87 | | | 51 | |

## 3. Results and Discussion

3.1 Dynamic viscosity of PVP solution in non-hydrate runs

This section presents the dynamic viscosity of PVP40 at 7000 ppm and high pressures, but at driving forces low enough not to trigger nucleation. Each datapoint in Fig. 3 is collected at a shear rate of 400 $s^{-1}$, averaging approximately 110 readings within a period of 10 minutes where the viscosities were stable at ±0.05 mPa·s after the perturbation due to a pressure increase.

As shown in Fig. 3, the aqueous solution exhibits a slightly decreasing trendline with respect to pressure, for all tested temperatures with the exception of 10 °C. The negative slopes from 0 to 8 °C conformed to the behavior of water viscosity at pressures from 0 up to 100 MPag.[19, 20] While the free-volume theory suggests that the viscosity of most liquids should increase as compression reduces free volume and increases resistance to flow, water at temperatures lower than 33.5 °C is an exception.[19, 21, 22] Aside from changes in free volume and thermal expansion coefficients, other structural organizations at the molecular level impact viscosity at low temperatures and pressures.[20] The addition of 7000 ppm PVP40 did not completely disrupt this unusual behavior, but some differences compared to pure water were still present.

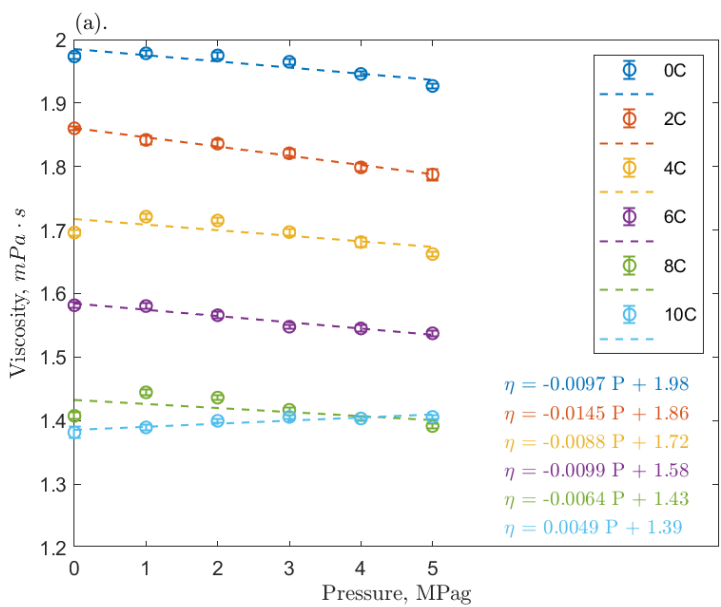

Fig. 3 Dynamic viscosities of PVP40 7000 ppm aqueous solutions as function of methane pressure with 95% confidence intervals and linear trendlines

Notably, the viscosity of the aqueous solution increased by a factor of $1.14 \pm 0.05$ compared to the water reference values.[17] This is due to the stronger frictional forces between the coiled or swollen polymer and solvent molecules, which contributed to a higher intrinsic viscosity of the solution.[23] Additionally, the dynamic viscosities at 4 and 8 °C, although decreasing overall with respect to pressure, performed a positive step change when pressurized from 0 to 1 MPag. The former is likely related to water's anomalous characteristics at 4 °C, the temperature at which its smallest molar volume was identified.[24] Two factors are disputing over the solution's viscosity evolution with respect to pressure: water's atypical structural properties and change in free volume theory. The step change observed at 8 °C represents the transition phase, beyond which the swollen polymer creates sufficient hinderance to water molecules' structure, such that the component of free volume gains more influence and results in positive slope for viscosity at 10 °C. The modified molecular interactions due to additives could potentially be factors that impact the effectiveness of PVP in disrupting hydrate cage structures.

3.2 Effect of PVP concentration on hydrate formation and growth

Methane hydrate formation from solutions of PVP40 at 700 ppm (Fig. 4) and 7000 ppm (Fig. 6) at 400 s$^{-1}$ was measured by the MCR302 as temporal evolutions of the dynamic viscosity of the hydrate slurries. At this shear rate, the torque limit imposed on the measuring system stops the measurements at approximately 1200 mPa·s, to protect the equipment from damage caused by the sample's rapid solidification.

Concerning the effect of PVP concentration on hydrate growth, Posteraro et al. (2015)[13] fit a sigmoidal curve to data points of methane consumption rate (µmol/s) with concentrations of PVP10 from 0.7 to 20,000 ppm in a static reactor. Whereas methane intake decreased drastically from 10 to 100 ppm, the curve reached a plateau for concentrations at and above 700 ppm. In this study however, the introduction of stirring resulted in a stronger correlation between initial PVP loadings and hydrate agglomeration (Fig. 9).

From Fig. 4, the addition of PVP (right) produced cleaner, more orderly viscosity profiles compared to the pure water panels (left). The presence of high-viscosity hydrate slurries, notably at around 500 mPa·s, visibly diminished, and the duration of low-viscosity slurries extended for longer periods. Averaging 84 hydrate-forming runs (87 minus 3 runs where nucleation did not occur within the experimental timespan), the time required for methane hydrates to reach 50 mPa·s nearly doubled in the presence of PVP40 700 ppm (1.9 times the water reference value). Although this system initially demonstrated beneficial effects for reducing growth, the addition of PVP40 700 ppm accelerated hydrate agglomeration in several cases. An example is illustrated by Fig. 5. The relative time $t_{PVP40\ 700\ ppm}/t_{water}$ to reach 200 mPa·s rapidly decreased to 1.2, with the lower error bar intersecting a ratio of 1.

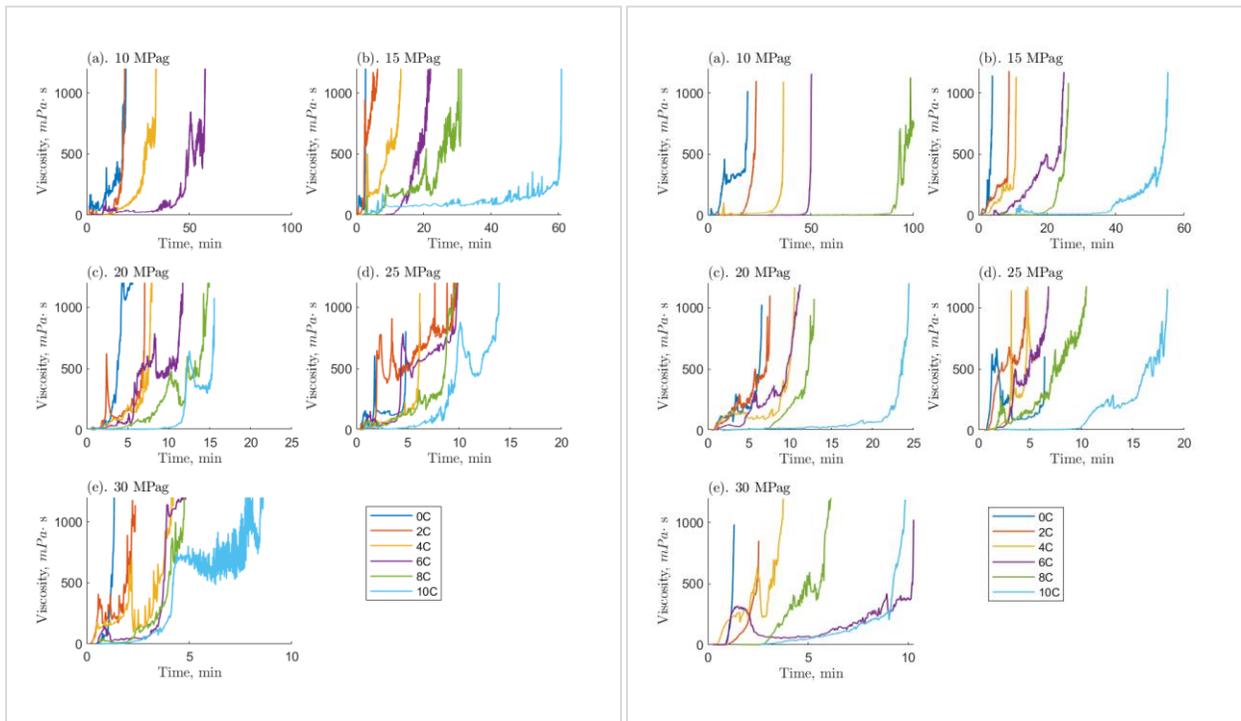

Fig. 4 Viscosity profiles of methane hydrate slurries at 400 s$^{-1}$ with respect to time for all test conditions with Left: reference system with pure RO water, adapted from Guerra et al. (2022)[17] and Right: PVP40 700 ppm

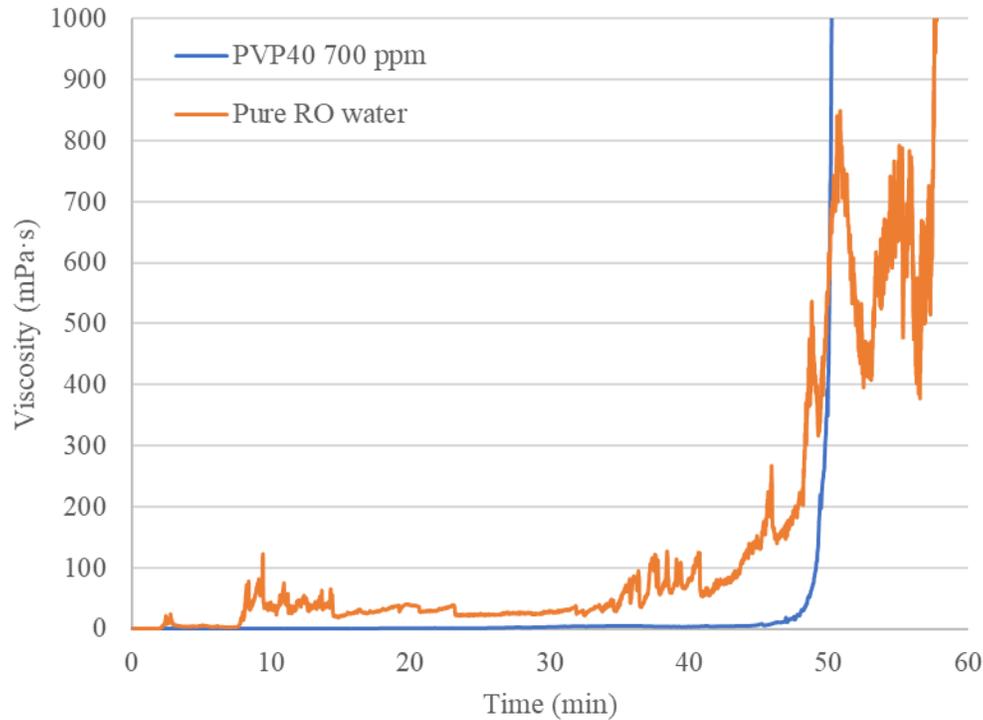

Fig. 5 Example of viscosity profiles with and without addition of PVP, figuring the case of PVP40 700 ppm 6 °C 10 MPag

At higher KHI concentrations, all viscosity curves behave with similar trends. The properties of PVP that suppress hydrate growth, followed by abrupt complete agglomeration, are more clearly illustrated in Fig. 6. At higher concentrations, the low-viscosity region extended to a significantly longer period. The time required for slurries to grow to 50 mPa·s with 7000 ppm increased to 8.2 times the reference value. The absence of high-viscosity hydrate slurries, as well as the steeper increase when inhibitory effects wore off, occurred in nearly all experimental conditions, with one exception at 6 °C 15 MPag. This anomalous case can be caused by the first hydrate clusters adhering on the DG geometry walls instead of growing in the bulk. Fig. 6 has been zoomed in to a y-axis limit of 500 mPa·s to provide more details of the low-viscosity region. In reality, the nearly vertical curves all extend to the torque limit.

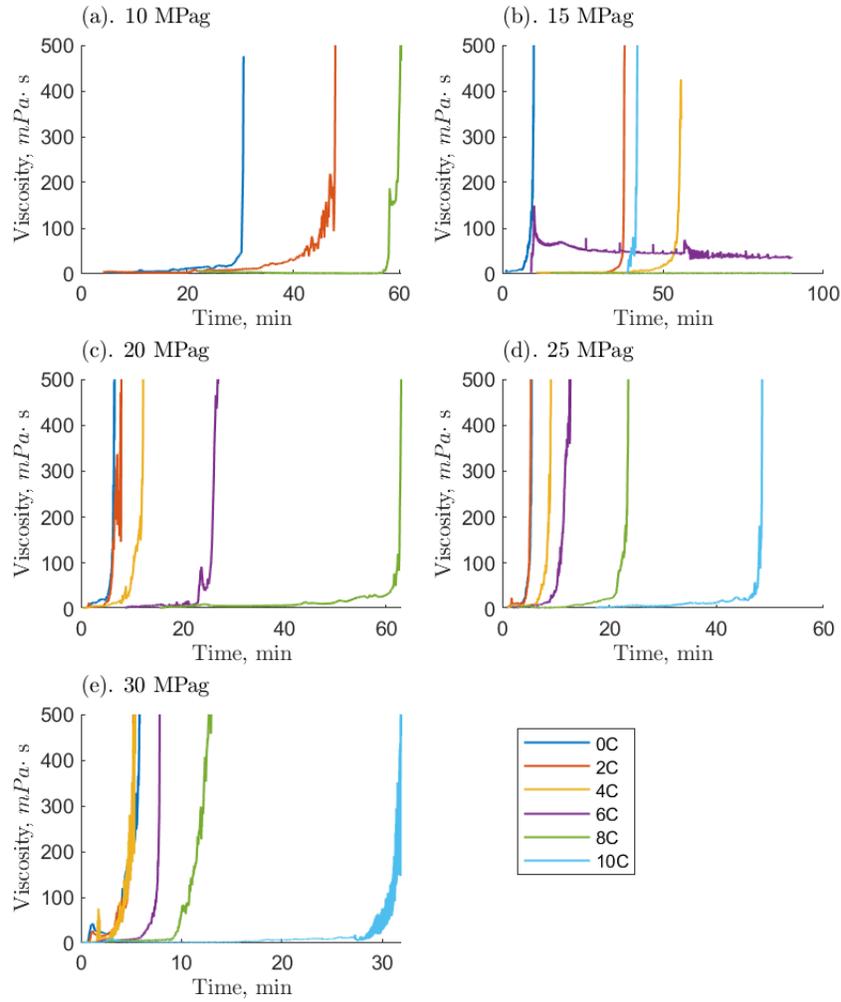

Fig. 6 Viscosity profiles of methane hydrate slurries at 400 s$^{-1}$ with respect to time for all test conditions with PVP40 7000 ppm

Fig. 7 provides a general idea of the difference in hydrate growth timescales between the two concentrations, presenting only the median values of the 3 runs to reach 200 mPa·s at each test condition where hydrates formed successfully, an issue discussed in detail later on. PVP40 7000 ppm was able to delay hydrate growth noticeably even at high-pressure driving forces, such as 20, 25, and 30 MPag. When analysing only the absolute time, it can be tempting to conclude that KHIs were most effective at low pressures and high temperatures. However, by compiling the results for relative time, no correlation was discovered between KHI's performance on growth rate and the tested temperature or pressure (Fig. 8). Therefore, all data were compiled in Fig. 9, presenting only the clear decreasing trend of relative time required for hydrate slurries to reach various viscosities.

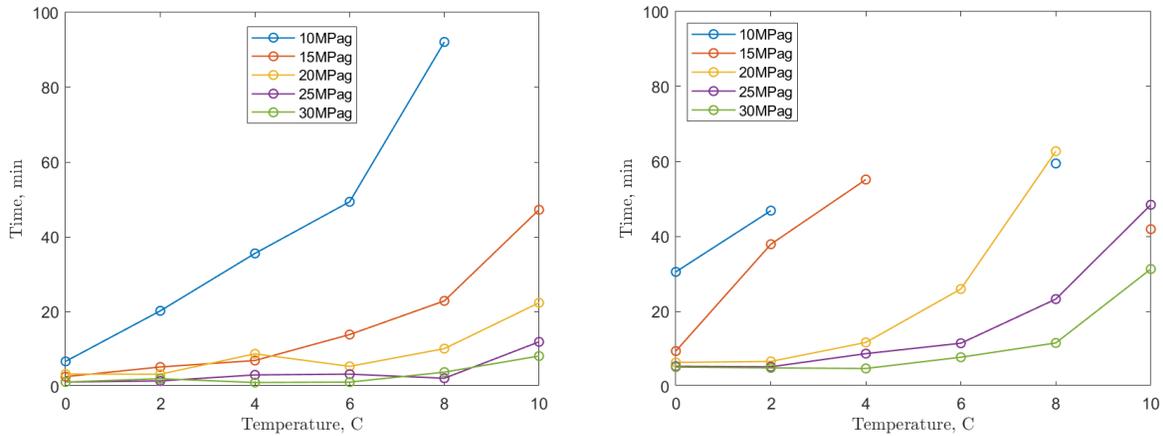

Fig. 7 Median time required for hydrate slurries to reach 200 mPa·s at all test conditions, for Left: PVP40 700 ppm and Right: PVP40 7000 ppm

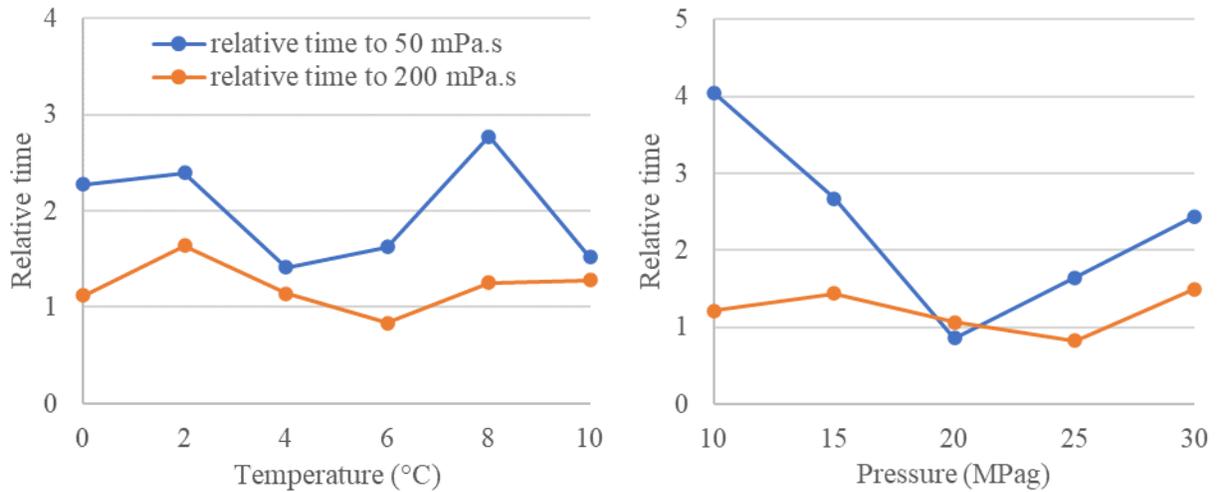

Fig. 8 Geometric mean of relative time $t_{PVP40\ 700\ ppm}/t_{water}$ to reach 50 and 200 mPa·s as function of temperature (left) and pressure (right)

The difference between the performance of the 700 and 7000 ppm solutions in reducing hydrate accumulation in the system was considerable, with 7000 ppm being more than 4 times more effective at the beginning of the growth phase. In addition, the mean agglomeration time for 700 ppm runs showed nearly no distinction compared to the water reference value, with a relative time of 1.04 to reach the torque limit.

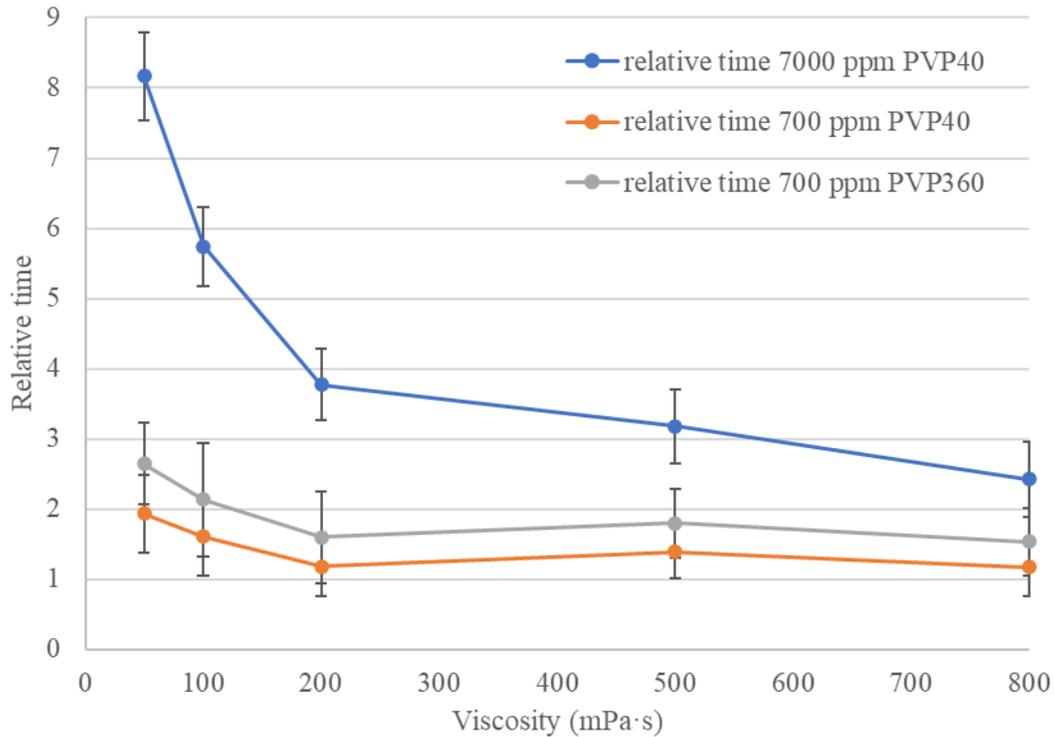

Fig. 9 Geometric mean of relative time $t_{PVP}/t_{water}$, ratio of total time required for the hydrate slurries to reach various viscosities for PVP40 700 ppm, PVP40 7000 ppm and PVP360 700 ppm, averaged over all test conditions, with 95% CI

The decoupling of methane consumption and agglomeration time in the presence of KHIs has also been observed by Sharifi et al. (2014),[25] with experiments performed in a different Anton Paar rheometer. While the cumulative gas consumption and initial viscosity decreased with the addition of KHIs, PVP favored the adhesion of suspended solids and caused a sudden rise in viscosity sooner than the control. This undesired behaviour, very similar to that manifested in the current study, was only observed in aqueous solutions. The introduction of hydrocarbon (i.e., 33 vol.% *n*-heptane) modified the molecular interactions and mass-transfer substantially, and the suspension remained dispersed much longer with KHIs compared to the control.

This decoupling of the two measured variables, gas intake and viscosity, can be attributed to two factors. One, it is associated with the depletion rate of PVP. As polymer molecules adhere onto hydrate surfaces, their concentration in the bulk liquid decreases.[26] Moreover, the diminution is not linear over time: the PVP uptake rate is highest at the beginning of the growth stage. This decrease is much more drastic in terms of percentage when the initial PVP loading is lower. While the initial hydrate particles might be stabilized regardless of the PVP content in the aqueous phase, in the later stage where mass-transfer plays a larger role in the inhibition mechanism, the system with lower initial PVP loading becomes less efficient in prolonging the desired slow growth. Additionally, a dynamic system continually creates new gas-liquid interfaces, hence why PVP's inhibitory performance cannot match that measured in static reactors, and quickly deteriorates with time.[10]

The second rationale relates to KHIs' potential to hinder interactions between water molecules to consolidate cage structures, disrupting nuclei before they reach critical size, thus prolonging the induction time.[8, 14, 15] While methane-water hydrates without additives began growing nearly immediately upon pressurization, with an average of 0.82 ± 0.67 minutes for induction time, nucleation phase with KHI addition was far more unpredictable and ambiguous.

The anti-nucleator characteristic of KHIs usually becomes more prominent at low pressures and high temperatures, but not always. For example, when analysing Table 2, one can notice that nucleation occurred within 90 minutes only 1 out of 3 times with 10 °C 20 MPag, but 2/3 times with 8 and 10 °C 15 MPag, despite the latter conditions exerting much lower driving forces. It is inconclusive as to whether interactions on the molecular level, which modified the water viscosity behavior as discussed in the non-hydrate measurements section, coincide with the accentuated anti-nucleator effect at 4 and 10 °C.

Table 2 Time required to trigger hydrate growth, for all test conditions with PVP40 7000 ppm

| Temperature (°C) | Time to hydrate growth (min) Pressure (MPag) | | | | | Nucleation: |
|---|---|---|---|---|---|---|
| | 10 | 15 | 20 | 25 | 30 | |
| 0 | 4.387 | 0.867 | 3.600 | 1.812 | 0.693 | |
| | -- | 53.76 | 0.840 | 0.008 | 0.002 | 0/3 |
| | -- | 1.733 | 0.880 | 0.933 | 0.640 | |
| 2 | 4.200 | 31.64 | 2.187 | 2.658 | 1.725 | |
| | -- | 36.31 | 2.333 | 0.002 | 0.002 | 1/3 |
| | -- | 1.960 | 1.096 | 0.960 | 0.747 | |
| 4 | 80.64 | 10.640 | 3.050 | 0.987 | 0.008 | |
| | -- | -- | 2.267 | 0.008 | 0.773 | 2/3 |
| | -- | -- | 1.493 | 1.760 | 0.693 | |
| 6 | -- | 8.817 | 8.680 | 2.000 | 5.195 | |
| | -- | -- | 3.100 | 2.567 | 0.002 | 3/3 |
| | -- | -- | 2.053 | 0.053 | 0.053 | |
| 8 | 22.12 | 14.78 | 15.49 | 7.977 | 5.687 | |
| | -- | 86.80 | 4.650 | 2.967 | 0.002 | |
| | -- | -- | 7.560 | 2.347 | 1.787 | |
| 10 | -- | 38.73 | 48.25 | 15.55 | 0.008 | |
| | -- | 29.59 | -- | 11.98 | 2.853 | |
| | -- | -- | -- | 2.640 | 2.640 | |

Increase in PVP concentration is another factor accentuating the anti-nucleator effect.[15] Indeed, as mentioned previously, only 3 runs amongst all tests with 700 ppm extended nucleation to over 90 minutes. Those took place at 2, 4 and 8 °C at 10 MPag, the lowest pressure. When comparing to the 7000 ppm table, the difference is significant in terms of both the number of runs and the highest affected driving force.

With regards to reproducibility, the induction time values were vastly different within the 3 runs at a single condition. For instance, as listed in Table 2 for the 2 °C 15 Mpag case, time required to enter the growth phase can vary between immediately after pressurization, just like its water counterpart, or after more than 30 minutes, despite the same system temperature and pressure. Although they are generally functions of supercooling or supersaturation of gas molecules in water, the times remain stochastic,[10, 12] and the number of repeated experiments per condition in the current report is insufficient to draw a clear trend.

Nuclei predominantly form at the gas-water interface, where the degree of supersaturation is highest. Then, hydrate particles get distributed into the bulk by mixing. However, it has also been recorded that the system can wait until its entirety becomes supersaturated before nucleation triggers uniformly in the bulk.[12] Various crystallization mechanisms are possible and are up to dispute, and certain might be more favourable for PVP's inhibitory effect. If nucleation was to occur in bulk, greater amount of dissolved PVP can engage in inhibition compared to if particles only form at the gas-liquid interface. Another peculiarity at high PVP concentration was that the dynamic viscosities during the nucleation phase were subject to unusual variations. There were very small fluctuations in the system, often less than 1 mPa·s, that are illustrated in Fig. 10. They either represent traces of nuclei that failed to reach the critical radius, or crystals of very small sizes that were successfully stabilized by the high amount of PVP and never further aggregated within the test period.

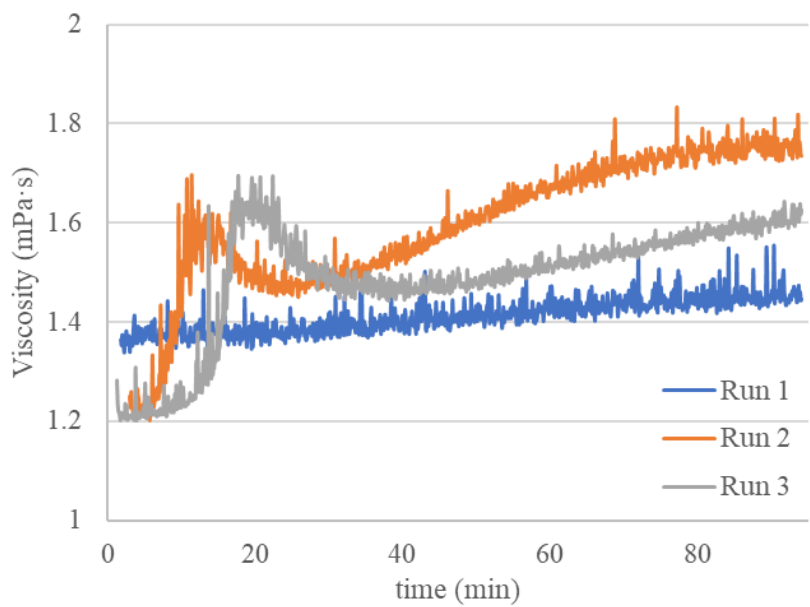

Fig. 10 Viscosity profiles of PVP40 7000 ppm at 6 °C 10 MPag starting from pressurization, potentially showing signs of nuclei under critical size, not considered as hydrate formation

3.4 Effect of PVP molecular weight

The methane consumption rates at concentrations from 0.7 to 20,000 ppm, tested using commercial PVP at 10, 40, and 360 kDa at 277.1 K and 6,282 kPa, have been reported to be uncorrelated to molecular weight.[13] When considering Fig. 11, however, the occurrence of high-viscosity slurries further diminished compared to PVP40 in the 10 MPag and 15 MPag panels in Fig. 4 (right), and the median agglomeration times are overall delayed, analogous to the high concentration set.

Since no correlation was found between inhibitory performance and pressure, 25 and 30 MPag were removed from the test conditions for the remaining experiment sets. When compiling all the hydrate growth runs (51 runs, of which 6 did not exhibit growth within the test period), the relative time was overall slightly higher than that of PVP40 700 ppm, with $t_{PVP360\ 700\ ppm}/t_{water}$ of 2.6 to reach 50 mPa·s. This ratio difference is, similar to the effect of concentration, most significant at 50 mPa·s and decreases at higher viscosities.

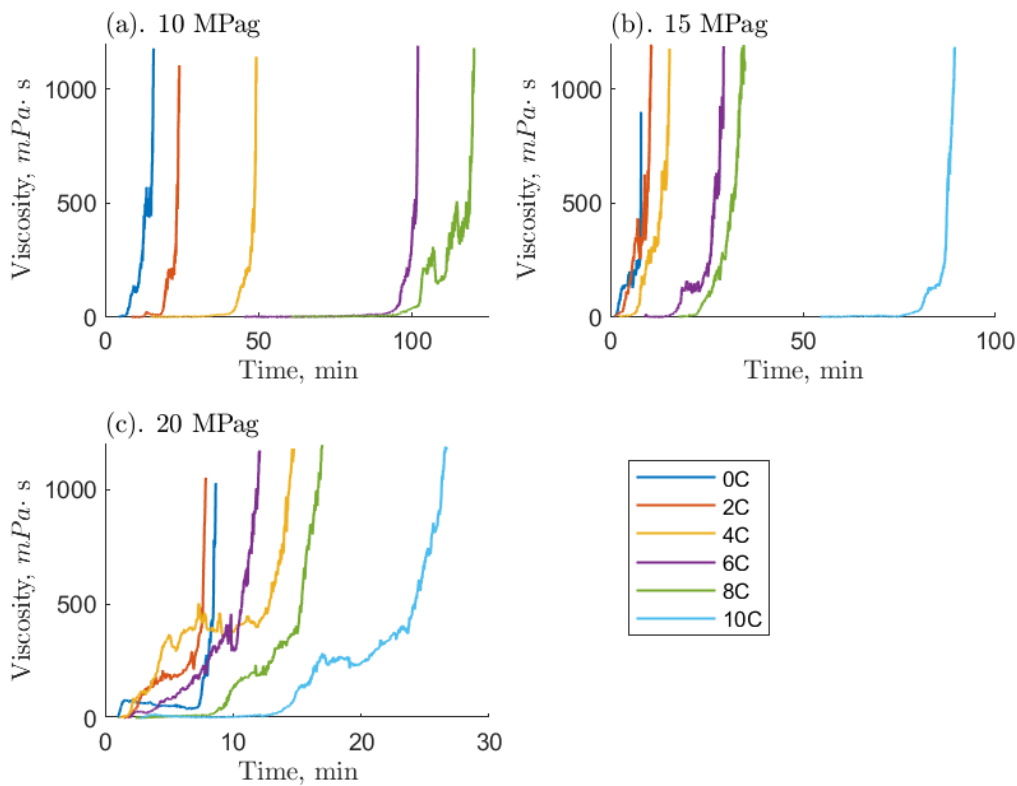

Fig. 11 Viscosity profiles of methane hydrate slurries at 400 s$^{-1}$ with respect to time for all test conditions with PVP360 700 ppm

The two aforementioned factors affecting hydrate growth: disruption of hydrate structures by PVP-water interactions during nucleation,[15, 27] and sharp PVP depletion rate in the bulk liquid at early growth stage,[26] can also be considered to explain the difference in performance between molecular weights. First, PVP360 contributes to a greater effective volume per unit mass, which is associated with the increase in intrinsic viscosity with respect to chain length.[23, 28] Hence, it is likely that more surrounding solvent molecules are affected, and the swollen polymer molecules create more structural interference to the water cages.

Ivall et al. (2015)[26] proposed a schematic of the early and late hydrate growth stages of PVP-inhibited systems with different initial loadings, where hydrate particles were of the same size at the beginning of growth, but since more available PVP chains occupied a larger surface area initially, particles with higher initial loading remained smaller for a longer period of time. An analogy can be drawn between the early stage of higher concentration and that of higher molecular weight. One polymer molecule at 360 kDa would be equivalent to multiple chains of PVP40 all adsorbed onto the surface of a single hydrate particle. Additionally, in simulations, longer PVP chains demonstrated better affinity with the methane hydrate surface due to more available binding sites.[29] Therefore, the inhibitory effect of PVP360 is greater at the early growth stage. However, this is accompanied by a more acute PVP uptake in the liquid. So, once surplus hydrate surfaces were created, the adhesion is prompter compared to shorter PVP chains.

3.5 Effect of shear rate on water system

At a lower shear rate of 80 s$^{-1}$, the measuring system was allowed to run for an extended time during rapid hydrate growth without damaging the equipment. Dynamic viscosities up to 4000 mPa·s were able to be recorded despite the torque limit, thus more information at the high-viscosity region became available.

It was hypothesized that a lower shear rate would increase the time required for hydrate formation (i.e., relative time $t_{80\ s^{-1}}/t_{400\ s^{-1}}$ greater than 1). According to Anton Paar,[30] shear rates above 300 s$^{-1}$ create eddy currents, and high velocity and agitation are factors that enhance hydrate formation, as observed in choke valves in pipelines due to the narrowing region.[1] Experimental data from Englezos et al. (1987)[12] also showed that induction period of methane hydrate was shorter for higher stir rates.

Amongst the 51 hydrate-forming runs, nucleation and growth were observed in only 38 runs. In other words, induction lasted longer than 90 minutes in 31% of the cases. Additionally, those runs had no clear trend compared to that presented in Table 2. While the PVP40 7000 ppm set had a weak correlation with pressure and temperature, the failed runs in the 80 $s^{-1}$ water set were scattered evenly across all conditions. On top of the nucleation rate, agitation speed also affects gas dissolution, as observed in $CO_2$ hydrate formation experiments in a gas-inducing agitating reactor by Li et al. (2016).[27] Taking into consideration the relatively small liquid-gas surface area the DG geometry provides (Fig. 1), the impact of shear rate seems to be significant on supersaturation and nucleation in this system.

Conversely, data of the 38 successful hydrate runs compiled into relative times $t_{80\ s^{-1}}/t_{400\ s^{-1}}$ of 0.75 to reach 50 mPa·s and 0.43 to reach 200 mPa·s, which is significantly faster. So, two opposing phenomena are occurring: on one hand, methane hydrates were less likely to form at 80 $s^{-1}$; on the other hand, 27 runs out of 38 exhibited a faster growth compared to pure water at 400 $s^{-1}$, averaging to a ratio much smaller than 1.

The latter phenomenon can also be justified by the nature of the DG measuring geometry. It was likely that a shear rate of 400 $s^{-1}$ was too high, so the hollow cylinder bob was able to blend and shred the hydrate slurries to a greater extent, decomposed clusters until rapid agglomeration eventually overcame the shear and stopped the spinning by triggering the safety torque limit. The faster growth at 80 $s^{-1}$ could be mainly related to the weaker physical deformation imposed on the sample. Therefore, shear rate has an effect on hydrate growth, but no effect was found on the ability of hydrates to form at certain driving forces.

In water-in-oil emulsions, steady state methane hydrate slurries behaved as a shear-thinning fluid from 1 to 500 $s^{-1}$. The slurries also exhibited a yield stress, which was greater at lower driving forces (lower pressure and higher temperature).[31, 32] This material property could explain why a three-step growth: 1) initial growth, 2) slurry phase, 3) agglomeration were observed at 400 s$^{-1}$.[17] Due to the shear-thinning property, slurries exist at higher viscosities when subjected to lower shear rate. From Fig. 12 and Fig. 13, a two-step growth specific to 80 s$^{-1}$ can be identified: 1) initial growth, and 2) extremely slow "growing" slurry in the high-viscosity region, typically above the limits of 400 s$^{-1}$ plots, which is thought to be attributed to strain and deformation rather than a decelerated adsorption rate of water and gas molecules on hydrate surfaces.

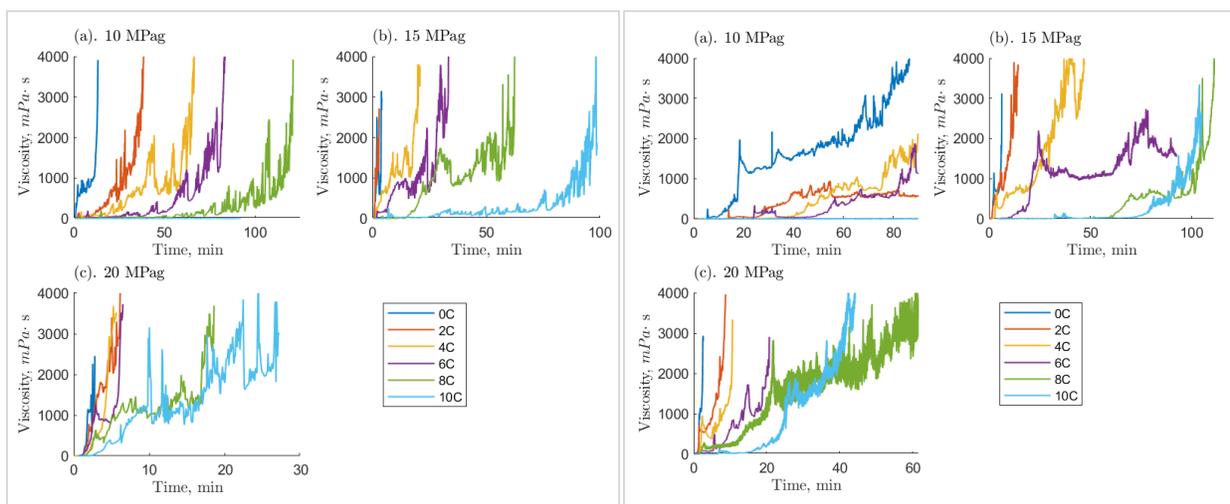

Fig. 12 Viscosity profiles of methane hydrate slurries at 80 s$^{-1}$ with respect to time for all test conditions with Left: pure RO water, and Right: PVP40 700 ppm

Contrary to Fig. 4, assembly at lower shear rates is greater with pure water rather than with addition of PVP. In the right panels, the presence of slurries is more prominent. The dissimilarity is most evident at the lowest pressure driving force. The trendline for 80 s$^{-1}$ in Fig. 13 differs from the 400 s$^{-1}$ curves in two ways: the relative time to reach 50 mPa·s is not the highest, and the familiar decreasing trend is only valid for viscosities below 800 mPa·s.

The former can be understood by inspecting the growth initiation step, as presented in the zoomed in Fig. 2 (right). In many cases, hydrate formation is signaled by a spike in viscosity, which then gradually re-equilibrates due to mixing because the first hydrate particles are distributed in the bulk. At higher shear rates, the first signals were weak, typically around ~5 mPa·s. At lower shear rate however, this signal increased by an order of magnitude, suddenly exceeding 50 mPa·s before rapidly falling back down (visible in Fig. 12 (Right)). The latter is more sensitive to noise, possibly because of the lower degree of mixing thus slower equilibration, so calculations of geometric average at low viscosities in Fig. 13 were compromised.

By performing a two-tailed distribution t-test with unequal variance on datasets from 50 to 800 mPa·s, $p < 0.05$ only for the datapoints at 200 and 800 mPa·s. While decreased shear rates has an effect on hydrate's overall viscosity growth, there is no sufficient evidence that this parameter affected PVP's performance. Whereas for 400 $s^{-1}$, slurries most often manifested around 200 mPa·s, they were more present at higher viscosities for shear rate of 80 $s^{-1}$.

Growth rates greatly decelerated at high viscosities with PVP, and there was more hydrate deformation causing the relative time to increase again when reaching 800 mPa·s: these effects may have been caused by the more porous morphology of hydrates with PVP.[16] Reduced methane intake at high water conversion may also be a factor in the system. However, considering the relatively short experimental timespans in this study, it is unlikely to be significant. For methane hydrate formation at 8-10 MPa, gas uptake only began plateauing after at least 75 minutes for water, and at least double that time with addition of 1 wt.% PVP90.[16, 33] It was also recorded that, while gas consumption with PVP remained lower initially, it eventually surpassed the water values after an extended period of time.[16] This provides further evidence that the effectiveness of PVP in reducing growth is strongest at the beginning of pressurization. Hence, the slow "growing" behavior with PVP is mostly attributed to hydrate morphology and its weaker yield strength. As the polymer disrupts growth by occupying a certain fraction of hydrate surfaces, onto which water and gas molecules can no longer adsorb, crystals with KHI addition have exhibited a different shape from those from pure water systems.[34] Therefore, the decreased mechanical strength of hydrates with PVP likely enhanced the extent of physical deformation in the DG geometry.[10]

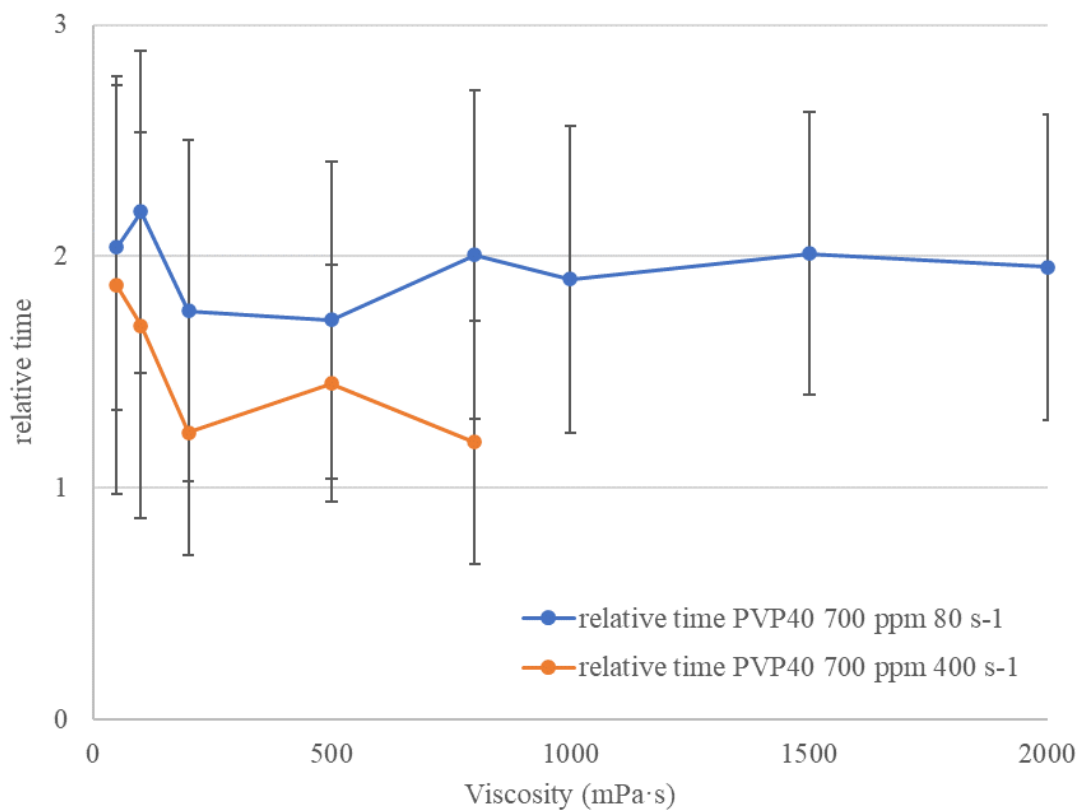

Fig. 13 Geometric mean of relative time $t_{PVP}/t_{water}$, ratio of total time required for the hydrate slurries to reach various viscosities for PVP40 700 ppm at 80 and 400 $s^{-1}$, averaged over test conditions of 10, 15 and 20 MPag, with 95% confidence intervals

## 4. Conclusion

Analysis of dynamic viscosity profiles of growing methane hydrate slurries in presence of PVP with different initial loadings, chain lengths, and shear rates provides new insights on their inhibitory performance and mechanism. At low concentrations, while PVP successfully reduced the occurrence of high-viscosity slurries, it did not prolong the total hydrate agglomeration time compared to the pure water system. Increasing PVP concentration in the aqueous phase significantly improves inhibitory effects in both the nucleation and growth phases, especially in a dynamic system, which exhibited vastly different behaviors compared to previous investigations where the growth rate was measured in terms of methane consumption.

At 7000 ppm of PVP40, high-viscosity hydrate slurries nearly completely disappeared, replaced by slow-growing crystals that typically remained under 50 mPa·s for the entire run until a sudden agglomeration. Dissolving PVP of greater molecular weight in water modified the hydrate viscosity profiles in an analogous manner as to the higher concentration solutions, but with a much weaker improvement in delaying agglomeration. On the other hand, both the amount and length of polymer chains are factors that potentially accelerate adhesion of hydrate particles, represented by the increasingly steeper viscosity rise at the end of each high-pressure test.

Decreasing shear rate from 400 to 80 s$^{-1}$ reduced the likelihood of hydrate nucleation. At the same time, methane hydrate grew much faster at a lower shear rate, a phenomenon attributed to the smaller physical deformation imposed on the shear-thinning samples. At 80 s$^{-1}$, the viscosity profiles of PVP40 700 ppm samples at 10 MPag were likely representative of the weaker yield strength of hydrate with addition of PVP, caused by their more porous morphology. Future works involve investigating the dynamic viscosity evolutions of methane hydrate slurries in presence of amphiphilic KHIs, which can engage in favourable molecular interactions with the gas phase, and potentially reduce occurrence of accelerated agglomeration as observed in this work.